\documentclass[seceq]{ptptex}

\usepackage[dvips]{graphicx}
\def\NPB{{\rm Nucl. Phys.} B}
\def\NPA{{\rm Nucl. Phys.} A}
\def\PLB{{\rm Phys. Lett.}  B}
\def\PRL{\rm Phys. Rev. Lett.}
\def\PRD{{\rm Phys. Rev.} D}
\def\PRC{{\rm Phys. Rev.} C}

\newcommand{\psla}{\mbox{\ooalign{\hfil/\hfil\crcr$p$}}}

\newcommand{\qsla}{\mbox{\ooalign{\hfil/\hfil\crcr$q$}}}

\newcommand{\vp}{\mbox{\boldmath $p$}}
\newcommand{\vq}{\mbox{\boldmath $q$}}

\newcommand{\get}{\mbox{${\tilde{g}}_e$}}
\newcommand{\ket}{\mbox{${\tilde{\kappa}}_e$}}

\newcommand{\epsi}{\mbox{$\varepsilon$}}

\newcommand{\tpsp}{\hspace{1.5em}}

\begin{document}

\title{\bf
In-Medium Effects in 
Eta Photo-production\\
through the $S_{11}$ Resonance \\
with the Relativistic Approach 
}

\author{Tomoyuki \textsc{Maruyama}$^{1,2,3}$
and Satoshi \textsc{Chiba}$^{3}$}

\inst{
$^{1}$ College of Bioresource Sciences,
Nihon University, Fujisawa, 252-8510, Japan \\
$^{2}$ Institute of Quantum Energy, Nihon University,
Tokyo, 101-8308, Japan \\
$^{3}$ Advanced Science Research Center, 
Japan Atomic Energy Research Institute,\\
Tokai 319-1195, Japan }

\recdate{January 28, 2003}

\maketitle

\begin{abstract}
In-medium effects in photon excitation of the $S_{11}$ resonance 
from a nucleon and its decay into 
an $\eta$-meson are studied
using the relativistic mean field approach.
We examine the in-medium effects by varying the Dirac fields of
the nucleon and the $S_{11}$ resonance separately.
We concluse that the Dirac fields of the nucleon, which
reduce the nucleon effective mass, enlarge the width of 
the $S_{11}$ resonance, and 
small Dirac fields of the resonance shift the peak position.
These combined effects can account for the in-medium properties observed in the experimental data
for $^{12}$C($\gamma$,$\eta$) reaction. 
\end{abstract}


\vfil
\eject

\newpage

\section{Introduction}

\tpsp
Photon induced reactions are useful for obtaining information concerning 
the deep interior region of nuclei because photons are distorted little 
in the incident channel.
Recent measurements of nuclear photo-absorption cross sections on several
nuclei heavier than $^{7}$Li \cite{Fro92} have shown clear evidence of
nuclear in-medium effects on isobar resonances: the resonance-like structures
caused by the D$_{13}$(1520) and F$_{15}$(1680)
excitations, clearly seen in the photo-absorption cross sections of
hydrogen\cite{Arm72a} and deuteron \cite{Arm72b}, disappear in heavier nuclei, 
while the peak of the P$_{33}$(1232) resonance still exists.

There have been several theoretical attempts to understand 
this disappearance \cite{Gia94,Alb94,Kon94,Eff97,Hir98}. 
At present, people are studiying the question of whether
these phenomena are caused by  the in-medium correction 
of the isobar resonances
such as broadening of the width \cite{Alb94}, or to other many-body processes
such as interference in two-pion photo-production processes 
in the nuclear medium \cite{Hir98}.
Regarding the isobar width in medium, Kondratyuk et al.\cite{Kon94} 
proposed that the collisional broadening plays a major role 
in the damping of the nucleon resonances in nuclei. 
However, Effenberger et al. \cite{Eff97} showed that such a large increase of 
the resonance width is not supported justified by an estimation of 
the collision broadening based on the $N-N^{*}$ interaction.

In order to solve this problem, recently Yorita et al. \cite{Yorita, Yorita-D}
carried out experiments on the ($\gamma$, $\eta$) reaction 
with a carbon target to study the $S_{11}$(1535) resonance.
This resonance couples strongly with the $N \eta$ channel and has 
a large branching ratio for $\eta$ decay.
Because the elementary $H(\gamma,\eta)H$ reaction below 1.0 GeV can be described 
in terms of the excitation of the $S_{11}$ resonance without 
any background process, 
it is reasonable to expect  only the $S_{11}$ resonance in ($\gamma,\eta$)
reactions with nuclear targets.
Furthermore, Yorita et al. \cite{Yorita, Yorita-D} analyzed 
their experimental data using the JAERI QMD code \cite{JQMD}. 
From the comparison between their experimental data and 
the calculation results, 
they found an evidence that the width of 
the $S_{11}$ resonance is broadened through the collisional process 
with nucleons \cite{Yorita}.
They were able to reproduce their data obtained using a C target 
with a resonance width
of 212 MeV (+ collisional broadening) while the elementary H($\gamma$,$\eta$)H
cross section is described better with a width of 154 MeV \cite{GRAAL}.

This QMD calculation includes effects of the final state interaction, such as
the absorption and multi-step collisions, and collisional broadening.
In this analysis the peak shift is the most clear evidence 
showing the discrepancy between the theoretical calculation and
the experimental data.
The peak position of the incident photon energy for 
the cross section obtained theoretically
is about 80 MeV lower than that observed experimentally if only the Fermi motion
of the nucleon is considered in the calculation.
The final state interaction from the absorption and the multi-step
collisional processes shifts the peak in the direction of higher energy by 
about 20 MeV to higher energy,
and the collisional broadening causes further shift of about 20 MeV.
Thus, the final peak-position is lower than the observed one 
by approximately 40 MeV.   
Therefore we need to consider other in-medium effects 
to account for the experimental results.

Bolkland and Sherif \cite{Sherif01} and 
Peters et al. \cite{Peters98} have calculated the exclusive
photo-production of $\eta$ meson with the DWIA using 
the relativistic mean field (RMF) approach \cite{Serot}.
However, because there are no exclusive data, unfortunately, 
they could not compare their results
with the experimental data and hence did not reach any meaningful conclusion.
However, their works provide a hint to understand the problem discussed above 
from the perspective of relativistic theory.
In the RMF approach, there are two kinds of Dirac mean fields, namely,
large attractive scalar and repulsive vector fields, which lead to
a small effective mass of the nucleon in the medium.
The variation of the Dirac mean fields changes the excitation energy and 
the width of the isobar resonance
because they depend on the effective mass of the nucleon, 
like the isobar resonance itself.

In this paper we investigate the in-medium effects in the $S_{11}$ resonance 
excitation from nucleon and its decay into an $\eta$-meson 
by introducing the mean fields of the resonance within the RMF approach.
Then, we can examine the in-medium effects by varying the Dirac fields of
the nucleon and the $S_{11}$ resonance separately.
As the first step, actual calculations are performed for nuclear matter, 
and we focus our investigation on the peak position of the 
cross section rather than the absolute values. 

In the next section we give the formulation of our model.
Because the interaction between isobars and nucleons is not completely clear,
we cannot carry out a complete microscopic calculation.  
For this reason, we propose 
a model that is partially microscopic and partially phenomenological.
In Sec. 3 we give some results, and we summarize our study in Sec.4.

\section{Formalism}

\tpsp
In this section we briefly explain our formalism.
Unfortunately, isobar resonances have not yet been scuccessfully
described with the standard field theoretical scheme.
Their selfenergies and propagators cannot be consistently obtained 
from certain Lagrangians in the relativistic framework.
All expressions have to be constructed phenomenologically.

On the other hand, 
phenomenological studies based on the Breit-Wigner formula \cite{Waler69} 
have succeeded in explaining various experimental results 
with regard to the meson production and absorption.
The Breit-Wigner formula is constructed under simple assumptions,
and it can be easily derived in the nonrelativistic framework,
but it is independent of the details of the interaction.
To analyze the experimental results, hence, we must construct 
a theoretical model that is consistent with the Breit-Wigner formula. 
Our approach explained in this section is formulated
according to this requirement.

First, we define  the coupling of the $S_{11}$ resonance, the nucleon 
and the photon in our model.
We calculate the photo-absorption cross section with the impulse approximation,
so that the calculational results are gauge-independent. 
Using the Coulomb gauge, then, we use the  simplest form  
of the interaction Lagrangian between nucleon, the $S_{11}$ resonancne and 
the photon as,
\begin{eqnarray}
 {\cal L}_{\rm elmag} &= &
\get {\bar \psi}_R \gamma_5 {\vec \gamma} \psi_{N} {\vec A}  +  {\rm h.c.}~, 
\label{Lag1}
\end{eqnarray}
where $\psi_N$, $\psi_R$ and $A^\mu$ are the nucleon, $S_{11}$ resonance,
and photon fields, respectively. 
In addition, {\get} is a coupling constant, which is determined 
from the experimental data.
As is well known, the final results are frame independent
under the Lorentz transformation,
if we consider only photo-absorption or photo-emission.
As discussed later, we choose the above interaction,
which is different from that in Ref. \citen{Sherif01,Peters98},
in order for our treatment to be consistent with conventional treatments
of the elementary process. 

Now we study  eta production in photo-absorption with the impulse approximation 
in terms of the RMF approach.
We define the initial photon momentum $q$ as
\begin{equation}
q = (Q;\vq) = (Q;0,0,Q) .
\end{equation}
The photo-absorption cross section is proportional to 
the transverse response function: 
\begin{equation}
\sigma_{abs} \propto \frac{1}{Q} \Pi_{T} .
\label{crabs}
\end{equation}
The transverse response function $\Pi_T$ is
defined as the imaginary part
of the transverse polarization function:
\begin{equation}
\Pi_{T} (q) = {\rm Im} \{ C_{11} (q) + C_{22} (q)\} .
\label{PiT}
\end{equation}
Hence, the polarization function for the excitation
between nucleon and the resonance is defined as
\begin{equation}
C_{i j} (q) = -i \int \frac{d^4 p}{(2 \pi)^4} {\rm Tr} \{
S_R (p+q) \gamma_5 \gamma_i S_N (p) \gamma_5 \gamma_j \} ~,
\label{Cpol}
\end{equation}
where, $S_{N}$ and $S_{R}$ are the nucleon and $S_{11}$ resonance propagators,
respectively.

The nucleon propagator $S_N$  involves the vacuum part and
the density-dependent part, which describe nucleons occupying states
below the Fermi energy of the system:
\begin{equation}
S_{N} (p) = (p_{\mu}^{*} \gamma^{\mu} + M_{N}^{*} ) 
\left\{ \frac{1}{p^{*2} - M_{N}^{*2} + i \delta } + \frac{i \pi}{E_N^*({\vp})} n(\vp) \delta (p_0 - \epsi_{N}(\vp)) \right\},
\label{Nprog}
\end{equation} 
where $n(\vp)$ is given by
$n(\vp) = \theta(p_F - |\vp|)$  with the Fermi momentum $p_F$.
In this expression, $M^*$ and $p_{\mu}^*$ are the effective mass
and the kinetic momentum, defined as
\begin{eqnarray}
M^{*}_N & = & M_N - U_s (N) , \\
p_{\mu}^{*} &=& p_{\mu} - U_{\mu} (N) ,
\end{eqnarray}
where $U_s$ and $U_\mu = U_0 \delta^{0}_{\mu}$ are the scalar and vector
mean fields.
In addition, the effective kinetic energy $E_{N}^{*} (\vp)$
and the single particle energy $\epsi_{N}(\vp)$ are defined
as
\begin{eqnarray}
E_{N}^{*} (\vp) &=& \sqrt{ \vp^2 + M_N^{*2} } , \\
\epsi_{N}(\vp) &=& E_{N}^{*} (\vp) + U_0 .
\end{eqnarray}

Similarly, we take the propagator of the $S_{11}$ resonance with scalar 
and vector fields $U_s (R)$ and $U_\mu (R) = U_0 (R) \delta^{0}_{\mu}$ to be 
\begin{equation}
S_{R} (p) = \frac{ p_{R\mu}^{*} \gamma^{\mu} + M^{*}_R }
 {p_{R}^{*2} - M^{*2}_R - i \sqrt{p_R^{*2}}\Gamma_{R} } 
\label{Rprog}
\end{equation} 
with
\begin{eqnarray}
M^{*}_{R} &=& M_R - U_s (R) , \\
p_{R \mu}^{*} &=& p_{\mu} - U_{\mu}(R) . 
\end{eqnarray}
The denominator of Eq. (\ref{Rprog}) is derived using the Breit-Wigner ansatz 
\cite{Eff97,Waler69,Peters98}.

Because the nucleon and meson that decay from the $S_{11}$ resonance
are coupled in the $s$-wave,
the energy dependence of the width of $S_{11}$ is determined
by the phase-space volume of decayed particles and the form factor of
the resonance. 
Its expression is given in Ref. \citen{Waler69} as
\begin{equation}
\Gamma_{R} = \Gamma_{\pi} + \Gamma_{\eta}
= \Gamma_{0} \left[ b_{\pi} \frac{k(\pi)}{k_{0}(\pi)} x_\pi + b_{\eta} \frac{k(\eta)}{k_{0}(\eta)} x_\eta \right]
\end{equation}
where $k(\pi)$ and $k(\eta)$ are momenta of the decayed pion and eta 
in the rest frame of the resonance,
$k_0(\pi)$ and $k_0(\eta)$ are the pion and eta momenta at the resonance pole
($s_R = M_R^2$),
$b_{\pi}$ and $b_{\eta}$ are the branching parameters for pion and eta channels, 
and $x_{\pi}$ and $x_{\eta}$ are the form factors given by
\begin{equation}
x_{\pi} =
\frac{k_0^2 (\pi) + c^2_{\pi}}{k^2 (\pi) + c^2_{\pi}}
~~~,~~~~
x_{\eta} =
\frac{k_0^2 (\eta) + c^2_{\eta}}{k^2 (\eta) + c^2_{\eta} },
\end{equation}
where  $c^2_{\pi}$ and $c^2_{\eta}$ are determined from experimental
data \cite{Krusche}.

The width must be modified by the Pauli-blocking effect of decayed nucleons 
as well as by the mean fields.
To include this effect, 
we define the width of the resonance in the medium 
as a function of the phase space.
When the nucleon with momentum $p$ absorbs the photon and is excited
to the resonance, the phase space volume of the decayed
particle becomes  
\begin{equation}
k^{2} (i) \rightarrow \langle k_{i}^{2} \rangle 
= \frac{1}{4 \pi} \int d^3 k \delta (k - k_N)
\label{decay-k1}
\end{equation}
where 
\begin{equation}
k^2_N = \frac{1}{4 s_{\rm eff}} \{(s_{\rm eff} - M_N^{*2} - m_i^2)^2 
                                    - 4s_{\rm eff}M_N^{*2} \} 
\end{equation}
with
\begin{equation}
s_{\rm eff} = (p^* + q )^2  .
\end{equation}
We transform the above integration in the rest frame of the resonance 
into the laboratory frame and add the Pauli-Blocking factor
$( 1 - n(\vp') )$.
Then we obtain 
\begin{equation}
< k_{i}^{2} > 
= \frac{\sqrt{s_{\rm eff}}}{2|\vp_R|}
k_N \int^{p_{+}}_{p_{-}} dp' \frac{p'}{E_N^*(\vp')} ( 1 - n(\vp') )
\label{decay-k}
\end{equation}
with
\begin{eqnarray}
&&\vp_R = \vp + \vq
\\
p_{\pm} &=& \frac{1}{ \sqrt{s_{\rm eff}} } 
| \{ E_N^*(\vp) |\vp_R| \pm (E_N^*(\vp) + Q) k_N \} | .
\end{eqnarray}

Within the impulse approximation, 
only contribution to the response function (\ref{PiT}) comes from 
the density-dependent part for the nucleon propagator (\ref{Nprog}) 
and the imaginary part of the resonance propagator (\ref{Rprog}):
\begin{eqnarray}
S_{N} (p) &\approx& \frac{i \pi}{E_N^*({\vp})} n(\vp) 
(p_{\mu}^{*} \gamma^{\mu} + M_{N}^{*} ) 
\delta (p_0 - \epsi_{N}(\vp)) .
\label{NprogD}
\\
 S_{R} (p) &\approx& - i( p_{R\mu}^{*} \gamma^{\mu} + M^{*}_R )
 \frac{\sqrt{p^{*2}_R} \Gamma_R }
{(s^*_R - M_R^{*2})^2 + p^{*2}_R \Gamma_R^{2} }.
\label{RprogI}
\end{eqnarray} 
Substituting these equations into Eq.(\ref{Cpol}), we obtain the transverse
response function $\Pi_T$.
When this is done, we must multiply $S_R$ by $\Gamma_{\eta}/\Gamma_R$ 
in Eq.(\ref{Cpol}) before integrating over the Fermi distribution
in order to account only for the eta decay of the resonance.

As shown in Eq.(\ref{crabs}), $\Pi_T/Q \rho_B$ is proportional to 
the contribution to the cross section from one nucleon.
Therefore, we define the arbitrary normalization factor ${\cal A}_{p}$ 
for a proton and ${\cal A}_{n}$ for a neutron and
obtain the eta photo-production cross section 
of the nuclear target with proton number $Z$ and neutron number $N$ 
\begin{equation}
\sigma_{\eta} =  ({\cal A}_{p} Z + {\cal A}_{n} N )
 \frac{4 {\cal F}_{abs}}{\rho_B Q}
\int \frac{d^3 \vp}{(2 \pi)^3} n( \vp ) W_{T}
{\sqrt s_R^*} f(s_R^*)\frac{\Gamma_{\eta}(s_{\rm eff})}{\Gamma_{R}(s_{\rm eff})} . 
\label{gamcrs}
\end{equation}
Here,
\begin{equation}
s_R^* = ( p + q^*)^2,
\label{srstar}
\end{equation} 
where $q^*$ is effective photon momentum defined as
\begin{equation}
q^* = (Q_0^*;0,0,Q ) = (Q+U_0(N)-U_0(R);0,0,Q)  ,
\end{equation}
and $W_T$ is defined as
\begin{equation}
W_T = W_{11} + W_{22} ,
\label{trsel}
\end{equation}
with
\begin{equation}
W_{\mu \nu} = \frac{1}{4 E_N^*(\vp) {\sqrt s_R^*}} 
{\rm Tr} \{ \gamma_{\mu} \gamma_5
(\psla^* + \qsla^* + M_R^*) \gamma_5 \gamma_{\nu} (\psla^* + M_N^*) \}.
\label{gamel}
\end{equation}
Here, the detailed expression of {$W_T$} in Eq. (\ref{gamcrs}) becomes
\begin{equation}
W_T = \frac{2}{E_N^*(\vp) {\sqrt s_R^*}} 
\{ E_N^{*2} (\vp) - p_z^2  + E_N^{*} (\vp) Q_{0}^{*}
- p_z Q + M^{*}_N M^{*}_{R} \} .
\label{WTr}
\end{equation}
In addition, the function $f(s^{*}_R)$ in Eq.(\ref{gamcrs}) is given with 
\begin{equation}
f(s^*_R) = \frac{\sqrt{s^*_R} \Gamma_R }
{(s^*_R - M_R^{*2})^2 + s^*_R \Gamma_R^{2} },
\label{mass-dis}
\end{equation}
where $\Gamma_{\eta}/\Gamma_{R} $ represents the decay ratio 
of the eta from the resonance, 
and  ${\cal F}_{abs}$ is the absorption factor for 
eta propagation, which is taken to be arbitrary in this work.

We determine the normalization factor
${\cal A}_p$ from the elementary cross section 
of the {$\gamma+p~{\rightarrow}~\eta$} reaction.
In our model, this cross section is obtained from Eq.(\ref{gamcrs}) as 
\begin{equation}
\sigma^{el}_1 = 2 {\cal A}_{p}
\frac{ M_N + M_R + Q }{Q} f(s_R) \frac{\Gamma_{\eta}}{\Gamma_{R}}  ,
\label{elm-crs1}
\end{equation}
where $s_R = (p+q)^2$ with $p = (M_N;0,0,0)$.
On the other hand,
this cross section is usually parametrized in 
the Breit-Wigner form \cite{Eff97,Krusche} as 
\begin{equation}
\sigma_{\eta}^{BW} = |A_{\frac{1}{2}}(p)|^2
\frac{M_N (M_R^2 - M_N^2)}{M_R^2 Q_c} \sqrt{s_R}  
f(s_R) \frac{\Gamma_{\eta}}{\Gamma_{R}} ,
\label{BW-paraC}
\end{equation}
where $Q_C$ is the photon momentum in the center-of-mass system
of the incident photon and proton.
We can rewrite this in the laboratory frame as
\begin{equation}
\sigma_{\eta}^{BW} = |A_{\frac{1}{2}}(p)|^2
\frac{M_R^2 - M_N^2}{M_R^2 Q} s_R  
f(s_R) \frac{\Gamma_{\eta}}{\Gamma_{R}} ,
\label{BW-para}
\end{equation}
using
\begin{equation}
Q_C = \frac{Q M_N}{\sqrt{s_R}}.
\end{equation} 

From the condition that $\sigma^{el}_1 = \sigma_{\eta}^{BW}$ at $s_R = M_R^2$
we can determine the factor ${\cal A}_p$ to be
\begin{equation}
{\cal A}_p = \frac{M_N (M_R - M_N)}{(M_R + M_N)} |A_{1\over2}(p)|^{2} 
\end{equation}
Furthermore, the relation between ${\cal A}_p$ and ${\cal A}_n$ is given 
in Refs. \citen{Eff97,Krusche} as
\begin{equation}
{\cal A}_n = \frac{2}{3} {\cal A}_p .
\end{equation}

Here we give a comment. 
$W_T$ exhibits a relativistic effect through the Lorentz transformation 
property of the Dirac spinor,
and therefore $W_T$ becomes constant in the nonrelativistic limit.
Such a nonrelativistic scheme actually gives the Breit-Wigner
form  \cite{Eff97,Krusche} 
for the cross section of the eta photo-production for  the proton
target.
In the center-of-mass system, in fact, the cross section 
$\sigma_{\eta}^{BW}$ is equivalent to our expression
when $W_T$ is taken to be $W_T = W_T^{NR} (CM) $  given as
\begin{equation}
W_T^{NR} (CM) = \frac{(M_R + M_N)^2}{M_R^2} .
\label{WTnrCM}
\end{equation}
In the laboratory frame, in addition, this $W_T$ becomes 
\begin{equation}
W_T^{NR} (L) = \frac{(M_R + M_N)^2 \sqrt{s_R}}{M_R^2 M_N} 
=  \frac{(M_R + M_N)^2}{M_R^2} \sqrt{ \frac{( M_N + 2 Q )}{M_N} } .
\label{WTnr}
\end{equation}

Here, we should give one more comment.
In Refs.\citen{Sherif01, Peters98}
a Lagrangian that differs from ours is used.
It is given by
\begin{equation}
{\cal L}_{\rm elmag} ^{(2)}= 
\ket {\bar \psi}_R \gamma_5 \sigma_{\mu \nu} \psi_{N} F^{\mu \nu} 
+ {\rm h.c.} ~.
\label{Lag2}
\end{equation}
Using this Lagrangian density,
the $p$($\gamma$,$\eta$) cross section is obtained in the same way as
\begin{equation}
\sigma^{el}_{2} =
\frac{ 4 M_N^2 Q }{M_R (M_R^2 - M_N^2)}  |A_{1\over2}(p)|^{2} 
f(s_R) \frac{\Gamma_{\eta}}{\Gamma_{R}}  .
\label{elm-crs2}
\end{equation}
The above cross section $\sigma^{el}_{2}$ differs
from $\sigma^{el}_{1}$ and $\sigma^{el}_{BW}$ by a factor of $Q^2$ 
in the limit $Q \rightarrow 0$, 
because $W_T$ is proportional to $Q^2$ in this limit.
Thus, this electromagnetic interaction is 
inconsistent with the Breit-Wigner formula, and therefore it cannot be used
for the analysis of experimental data.

Contrastingly,
the Lagrangian we chose has the simplest form that satisfies 
the $s$-wave coupling in the low-energy limit. 
Our expression can indeed reproduce the phenomenological
cross section and the experimental data for the elementary process 
as shown in Fig.1.
This implies that the Lagrangian we have chosen is reasonable for 
our purpose. 
Recently a similar electromagnetic interaction was derived
using a vector dominat model in Ref. \citen{Chiang}.
Our method must corresponds to  an approximation  
of this vector dominant approach. 

The numerical difference between {$\sigma^{el}_1$} and {$\sigma^{el}_2$} 
is presented in the next section.

\section{Results and Discussion}

\tpsp
In this section we reports the results of actual calculations.
The parameter values for the $S_{11}$ resonance are taken as
$M_R~=~1.540$ GeV, $\Gamma_0~=~0.15$ GeV, $b_{\eta}~=~0.55$,
and $c^2_{\pi}~=~c^2_{\eta} = 0.25$ GeV$^2$.

In Fig. 1 we plot our result ($\sigma^{el}_1$) with a solid curve, and 
the experimental data \cite{GRAAL,Krusche} are indicated by the filled circles
and squares.
Our result reproduces the experimental data quite well.
Furthermore, we plot the cross section $\sigma^{el}_2$ (\ref{elm-crs2}) 
obtained with the Lagrangian density ${\cal L}^{(2)}$ in Eq.(\ref{Lag2}) 
with the dashed curve, while the best-fit phenomenological expression
$\sigma_{\eta}^{BW}$ is represented by the dotted curve.
We can see that $\sigma^{el}_2$ does not agree with 
either $\sigma^{el}_1$ or $\sigma_{\eta}^{BW}$, at least when the same values of 
the resonance mass $M_R$ and width $\Gamma_0$ are used.

As commented above,
the difference between $\sigma^{el}_1$ and $\sigma_{\eta}^{BW}$
lies in $W_T$ defined in Eq. ({\ref{trsel}),
which is obtained from the Lorentz transformation property
of the Dirac spinor.
The agreement of two results implies that the relativistic effects, 
except for the kinematics, are not important in this elementary reaction.   

Next, we examine in-medium effects through the Dirac fields.
The parameter-sets for the nuclear selfenergies are taken from PM1 \cite{K-con}
(namely, $BE = 16$ MeV, $M^*/M=0.7$ at $\rho_0 = 0.17$fm$^{-3}$).
By contrast, we do not have any information 
on the $S_{11}$ resonance mean fields.
However, it is not unnatural to assume that the $S_{11}$ resonance potential 
is not large:
\begin{equation}
\frac{|U_{s}(R) - U_{0}(R)|}{M_R} \ll 1
\label{Rpot}
\end{equation}
Then, we fix the ratio of the nucleon and the $S_{11}$ resonance self-energies as
\begin{equation}
u_r = \frac{U_{s}(R)}{U_{s}(N)} = \frac{U_{0}(R)}{U_{0}(N)} ,
\label{ratio}
\end{equation}
and examine the effects of  the $S_{11}$ resonance mean fields
by varying the ratio $u_r$.
Of course, $U_s(R)$ and $U_0(R)$ can be varied independently, but
this variation is restricted to only a few tens of MeV by the condition 
(\ref{Rpot}), 
while the depth of the mean fields is about several hundred MeV.
Such small variation is not thought to affect the results of qualitative studies.

In the actual calculation we omit the Pauli blocking factor 
in Eq. (\ref{decay-k}),
because the photon momentum is very large, 
and this factor does not contribute to the results of the calculation.
In addition, the value of the normalization constant ${\cal F}_{abs}$ was 
determined to reproduce the experimental data at the peak.
Here, we take ${\cal F}_{abs} = 0.83$.
This value is not very far from 
the value obtained from the QMD approach 
(${\cal F}_{abs}=0.75$ around $Q = 850$MeV ) \cite{Yorita}.

In Fig. 2, we compare the results for the photo-production of $\eta$ with the 
experimental data in the case of $^{12}$C target.
The solid curve, long-dashed curve and dashed curve represent the results 
for $u_r=0$, $u_r=0.5$ and $u_r=1$, respectively.
We can see that the result agrees well with the experimental data 
in the case {$u_r=0$}. 

Now we give qualitative discussions of the peak photon energy.
The peak energy for the mass distribution function (\ref{mass-dis})
is given by $s_R^* = M_R^{*2}$.
When integrating over Fermi sea,
the actual peak energy
is close to the energy corresponding to the condition that
the absorbed photon excites a rest nucleon to the resonance with 
mass $M^*_R$, that is,
\begin{equation}
M^{*2}_R = (M_N^* + U_0 (N) - Q )^2 - Q^2
\label{peakQ}
\end{equation}
We define ${\bar Q}$ form the above condition as
\begin{equation}
{\bar Q} = \frac{M_R^{*2} - (M_N + U_{c}(N))^2}{2(M_N + U_{c}(N))} 
\end{equation}
with
\begin{equation}
U_c (N) = - U_{s} (N) + U_{0} (N) .
\end{equation}

When $U_{s,0}(R) = 0$, which corresponds to the case $u_r = 0$, we have 
\begin{eqnarray}
{\bar Q} &=& \frac{M_R^2 - (M_N + U_{c}(N))^2}{2(M_N + U_{c}(N))} 
\nonumber \\
&\approx& \frac{M_R^2 - M_N^2}{2M_N}
 - \frac{M_R^2 + M_N^2}{2M_N^2} U_{c}(N) .
\label{Qbar}
\end{eqnarray}
This $U_c (N)$ is almost identical to the central potential
of a nucleon in the nonrelativistic picture.
This value of ${\bar Q}$ is slightly larger than the real peak energy, 
because of the factor $1/Q$ appearing in the equation.
Substituting the values of the mean fields into the above Eq. (\ref{Qbar}),
we obtain that $\Delta {\bar Q} = {\bar Q} - {\bar Q}_0 = 120$ MeV,
while the actual peak shift is about 80 MeV.
This analysis thus overestimates the peak shift.    
When $M_N^* = M_N$ [$U_s(N)=0$] with the fixed Fermi energy, 
in addition, the vector field $U_0(N) = - 54$ (MeV), and we obtain 
$\Delta {\bar Q} = {\bar Q} - {\bar Q}_0 = 100$ MeV.
We thus conclude that the binding effect of nucleons 
partially accounts for the peak shift, but it is not perfect.

In Fig. 3 we display the results 
for $u_r = 0$  with $M^*_N/M_N = 0.7$ (solid curve) 
and  $M^*_N/M_N = 1$ (dotted-dashed curve).  
The peak shift at $M^*_N/M_N = 1$ is about 40 MeV, 
while that at  $M^*_N/M_N = 0.7$ is about 80 MeV. 
In the case $M^*_N/M_N = 1$, in addition, 
the width is smaller than that at  $M^*_N/M_N = 0.7$.
This result indicates that the in-medium correction of 
the nucleon mass also contributes to the broadening of the width.

To confirm the above claim, we consider the momentum-dependent width of 
the $S_{11}$ resonance in medium.
In Fig. 4, we display the width of the resonance, 
which is created from the rest nucleon,
$\Gamma_{\rm r}(\eta)=\Gamma_{\eta}/(b_\eta \Gamma_0)$.
The results are plotted as functions of
the incident photon energy for $u_r = 0$  with $M^*_N/M_N = 0.7$ (solid curve) 
and  $M^*_N/M_N = 1$ (dotted-dashed curve) in Fig. 4(a), and the same results are also
plotted as functions of $\sqrt{s_{\rm eff}}$ in Fig. 4(b).

The small effective mass of the nucleon increases the phase space
of the decayed particle and enlarges the width 
for small value of the effective invariant mass $s_{\rm eff}$ (see Fig.4(b)).
For the rest nucleon, however, the effective invariant mass
becomes
\begin{equation}
s_{\rm eff} = M_N^{*} ( M_N^{*} + 2 Q ) .
\end{equation} 
Thus the small effective mass also reduces the effective invariant mass
at the fixed energy of the incident photon. 
Therefore the width with a small nucleon effective mass 
is smaller in the low photon momentum region  
and larger in the high momentum region.
This effect also shifts the peak and broadens the actual width of the
mass distribution function of the resonance in the medium.

The fact that $u_r = 0$ is preferred,
is similar to the finding of Yorita's analysis given in Ref. \citen{Yorita}
that a small binding effect and the enhanced width of the resonance 
in medium can yield results that reproduce the experimental data.
 
Finally, we examine relativistic effects other than the kinematics 
by changing the method
to relate the elementary cross section with the in-medium one.
As mentioned before, 
$W_T$ exhibits a relativistic effect through the Lorentz transformation 
property of the Dirac spinor,
and $W_T$ becomes constant in the nonrelativistic limit.
Therefore we substitute $W_T^{NR}(L)$ of Eq. (\ref{WTnr}) into Eq. (\ref{gamcrs}) 
and integrate the latter over nucleon momenta below Fermi level.
This calculation corresponds to folding the phenomenological 
elementary cross section $\sigma_{\eta}^{BW}$.
We refer to the results obtained with this cross section as C2, while the results 
given above ($\sigma^{eta}$) are referred to as C1.

Furthermore, we consider the situation in which the rest nucleon
in the nuclear medium absorbs the photon with momentum $Q={\bar Q}$.
In this case, $W_T$ becomes $W_T^{\rm eff}$ defined as
\begin{equation}
W_T^{\rm eff} = \frac{(M^*_R + M^*_N)^2 \sqrt{s_{\rm eff}}}{2 M^{*2}_R M^{*}_N} .
\label{WTef}
\end{equation}
Then, we refer to the results of the calculation employing $W_T^{eff}$ instead of $W_T$ as C3.

In Fig. 5 the results are displayed.
There the solid,  
dashed and dotted curves represent the results 
calculated C1, C2 and C3, respectively.
The calculations were carried out using at $M^*_N/M_N = 0.7$ 
and with $u_r=0$ (Fig 5(a)) and with $u_r=1$ (Fig 5(b)).
While the three sets of the results are almost the same for 
$u_r=0$ (Fig 5(a)), C3 is slightly smaller than others (Fig. 5(a)).  
Contrastingly,
three results are almost the same for $u_r=1$ (Fig. 5(b)). 
As commented above,
this difference comes from  $W_T$ defined in Eq. ({\ref{trsel}),
which is obtained from the Lorentz transformation property
of the Dirac spinor.
This relativistic effect is, however, not so large; 
only a small effective mass contributes to the results.

As stated previously, the purpose of this paper is 
to qualitatively study the peak shift and broadening width 
in view of the kinematics,
and we are not presently concerned with analyzing the absolute value of 
the cross section.
The results obtained with the three kinds of calculations (C1, C2 and C3) 
have nearly same shapes, and their peak positions coincide.   
Only the peak of C2 with $u_r=0$ is shifted to a position
higher than that of the others by about 20 MeV, but this difference is not very significant.
In this respect, our calculation using Lagrangian 
Eq.(1) can be verified.

\section{Summary}

\tpsp
In this paper we investigated in-medium effects for eta photo-production.
We used the RMF approach and calculated the eta production cross section, 
varying the Dirac fields of the $S_{11}$ resonance.
We find that small Dirac fields of the $S_{11}$ resonance
gives the best results, though the nucleon fields are not small.
This result can be explained with the following mechanism.

The small Dirac-fields of the $S_{11}$ resonance shift the peak
to higher energy through binding effects,
and the small effective mass of the nucleon enlarges the width.
The deep resonance Dirac-fields also enlarge the width, but they
suppress the peak shift, and they cannot account for the experimental
data.
In other words  the difference between the mean fields of the nucleon 
and the resonance modifies the kinematical relation in-medium between them 
and changes the shape of the eta photo-production cross section.

When extending the elementary cross section to the in-medium case, 
there is a difference between our results and those obtained using 
the phenomenological elementary cross section that corresponds to
the nonrelativistic one.
This difference is caused by the small effective mass in the nuclear medium.
In any case, this effect only changes the absolute value 
and does not change our discussion on the peak and width of the cross section.

Of course, there are several ambiguities in our treatment, and for this reasons, 
we cannot give a perfectly quantitative conclusion at present.
Regardng the photo-absorption process, we must take account of
the collisional broadening of the resonance width in the future.
In addition, the final state interaction also contributes to
the peak shift; the eta absorption from the nucleon is energy-dependent,
and hence the absorption factor must also be energy dependent.
It is confirmed in Ref. \citen{Yorita} that this energy-dependence of
the absorption and the multi-step process shift the peak position toward 
higher energy,
though this final state interaction, together with the collisional broadening, 
is not sufficient to accout for the experimental data.
The effects we elucidated in this paper are supposed to be smaller 
in finite nuclei, and the collisional broadening and the final state
interaction must contribute to the peak shift.
It is expected that the cooperating effects of the Dirac-fields and the final state interaction
account for the peak shift and the width of the cross section. 
Thus, we need to calculate the cross section for finite
nuclei including realistic eta absorption and multi-step effects.
For this purpose, the RBUU approach should be useful \cite{RBUU,TOMO2}.
In such a work, we may consider in-medium corrections for eta absorption
and collisions between $S_{11}$ and nucleon.
In addition, we may take account of 
coherent effects for eta production \cite{Peters98,Schev}.

It has recently beenn suggestted that the $S_{11}$ resonance is 
the chiral partner of the nucleon \cite{Chiral}.
The development of this study may suggest other restrictions that can eliminate
for the ambiguities.
Furthermore, the in-medium modification of  $\eta$ properties 
has been suggested from this chiral partner model \cite{Chiral2}.
These effects may change the peak energy and width of the cross section.
We should take into account such in-medium corrections in the future.

We would like to give several further comments.
From this work, we learn that the peak position and the width
of the cross section are determined by in-medium modification  
of the kinematical relation between the isobar resonance and decayed particles.
We conclude that the $S_{11}$(1535) resonance does not have 
large in-medium effects.
On the other hand,  
we need to introduce more drastic effects to account for the experimental data
for D$_{13}$(1520) and F$_{15}$(1680).
There is a strong possibility that the in-medium correction of the $\rho$-meson is one effect 
that can be useful in this regard.

\bigskip
\noindent
\section*{Acknowledgments}

We would like thank Prof. J. Kasagi and Dr. H. Yamazaki for useful
discussion and providing us with experimental data.

\newpage

\vspace*{5mm}
\hspace{0.5cm}
{\includegraphics[scale=0.75]{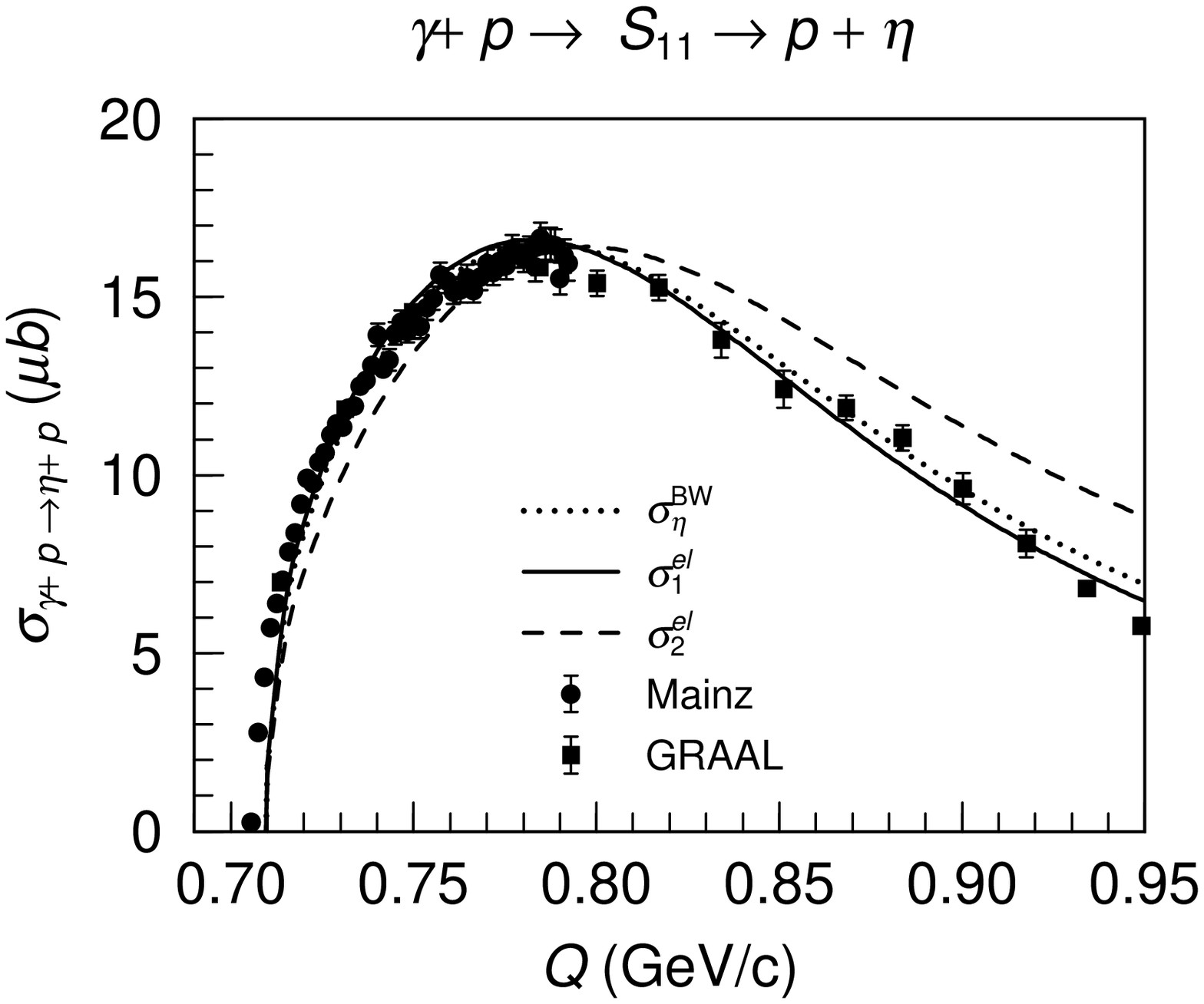}}

\bigskip
\noindent
{{\bf Fig.1}
Cross section for eta photo-production with proton targets.
The filled circles and squares represent the experimental data of Ref. \citen{Krusche}
and \citen{GRAAL}, respectively.
The solid, dashed and dotted curves represent the results of our calculation
based on $\sigma^{el}_1$ and $\sigma^{el}_2$ 
and of the phenomenological
parametrization $\sigma_{\eta}^{BW}$, respectively. 
}

\newpage

\vspace*{5mm}
{\includegraphics[scale=0.65,angle=270]{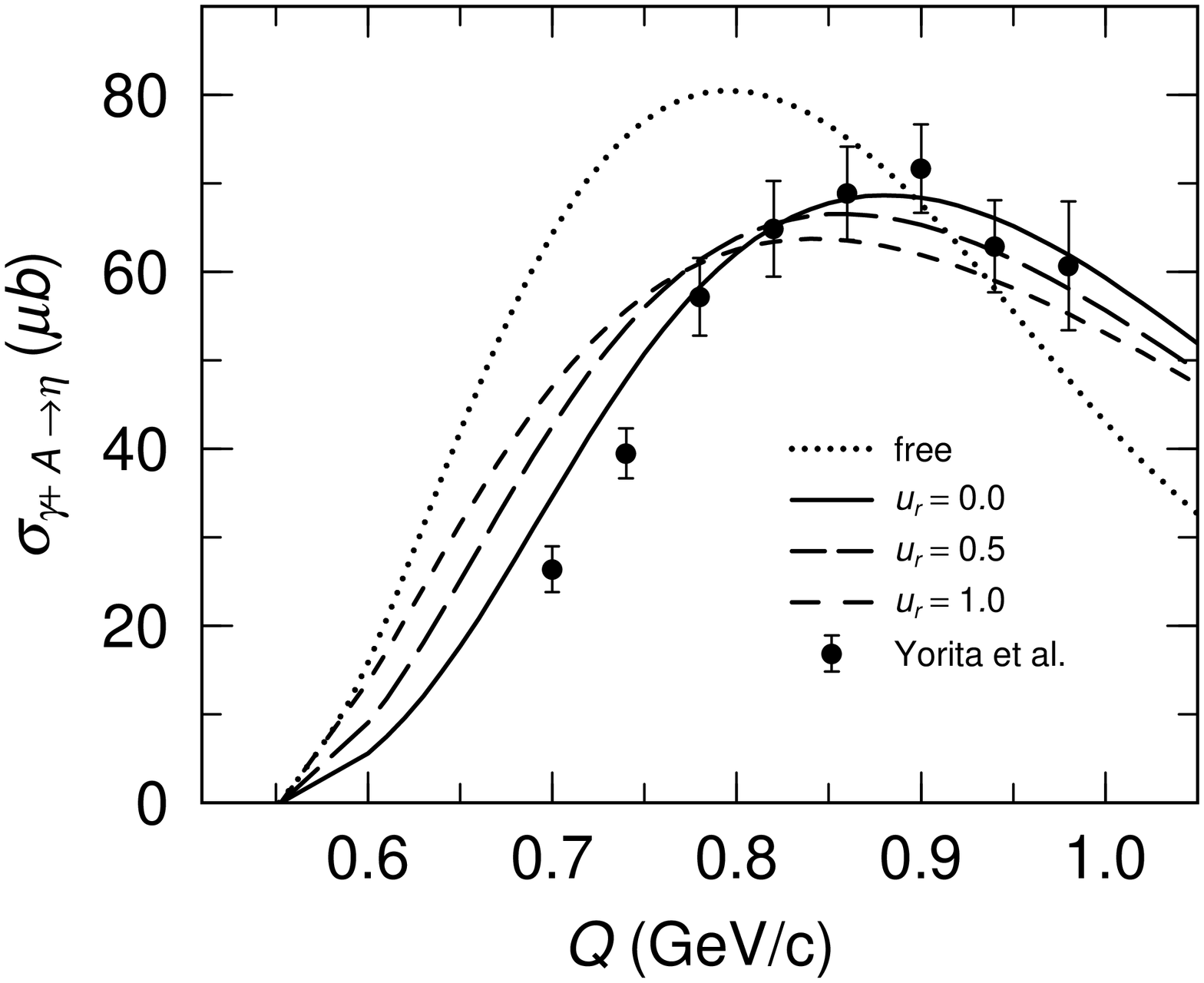}}

\bigskip
\noindent
{{\bf Fig.2}
Cross section for eta photo-production with $^{12}$C targets.
The solid, long-dashed and dashed curves represent
the results with $u_r = 0.0, 0.5$ and $1.0$, respectively.
The nucleon effective mass is taken to be  $M^{*}_N / M_N = $0.7.
The dotted curve represents the results obtained with a free Fermi gas.
The filled circles show the experimental data \cite{Yorita}.
}

\newpage

\vspace*{5mm}
{\includegraphics[scale=0.65,angle=270]{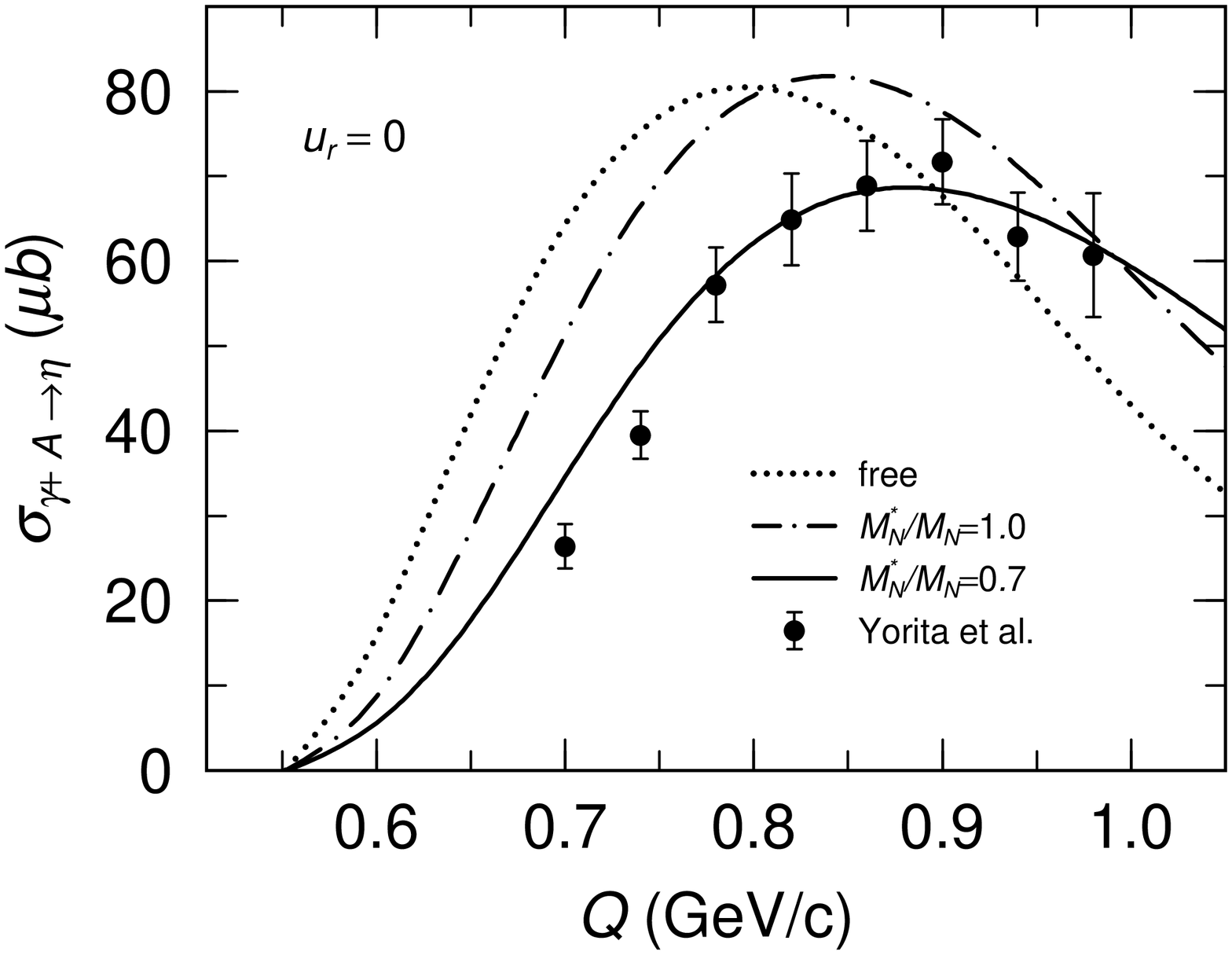}}

\bigskip
\noindent
{{\bf Fig. 3}
Cross section for eta photo-production with $^{12}$C targets.
The solid  and dash-dotted curves represent
the results obtained with nucleon effective masses 
$M^{*}_N / M_N = $0.7 and 1.0, respectively.
The dotted curve represents the results with a free Fermi gas.
No mean fields for $S_{11}$ resonances were considered, i.e.,
$u_r =0$. 
The filled circles show the experimental data \cite{Yorita}.
}

\newpage

\vspace*{5mm}
\hspace{1.cm}
{\includegraphics[scale=0.8]{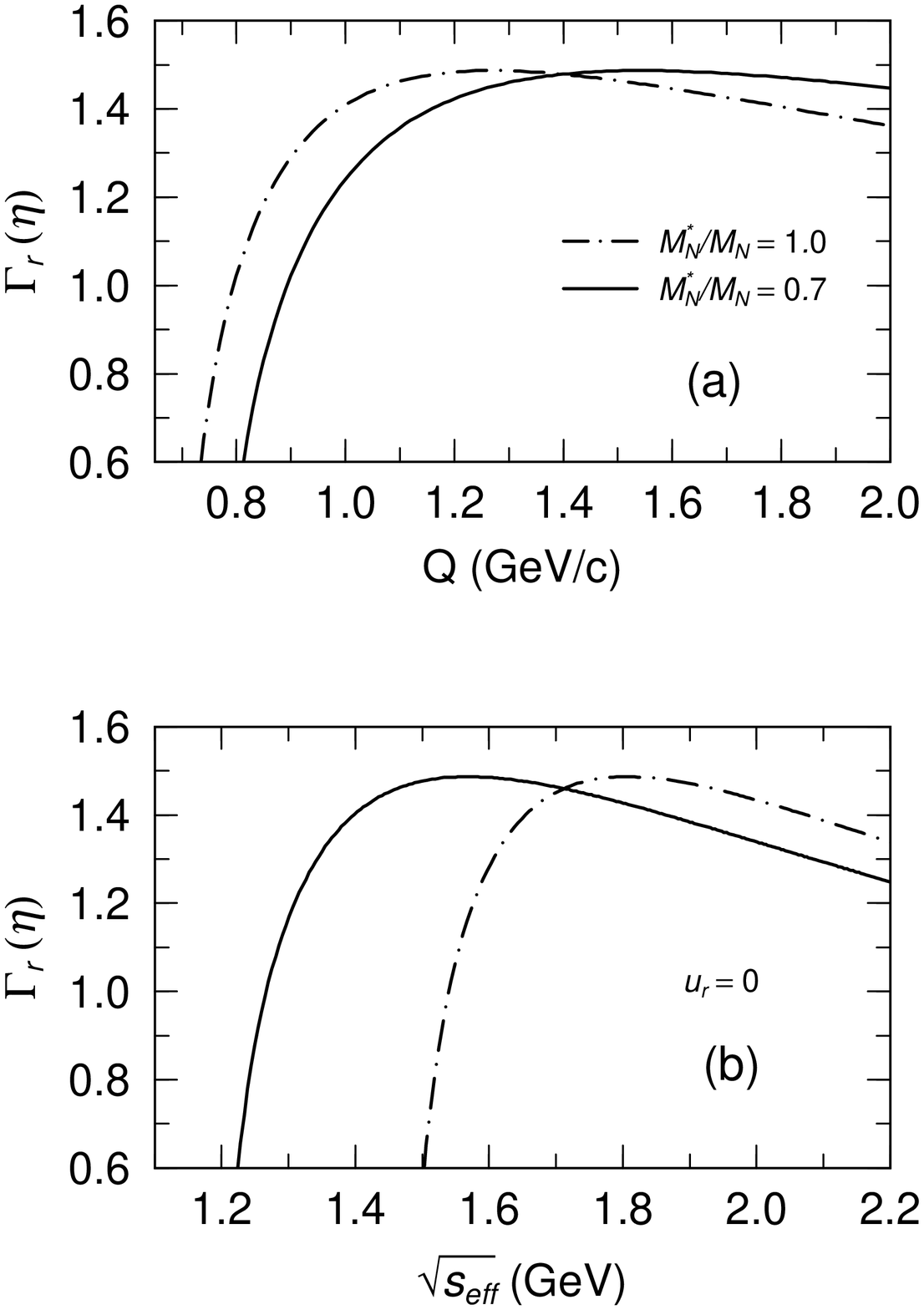}}

\bigskip
\noindent
{{\bf Fig.4}
The momentum dependence of the resonance width for $u_r = 0$
is plotted as a function (a) of the photon momentum and (b) 
of the effective invariant mass (b).
The solid and dash-dotted curves represent the results with 
$M^*_N/M_N$ = 0.7 and = 1.0,
respectively.
The details are explained in the text.
}

\newpage

\vspace*{5mm}
{\includegraphics[scale=0.8]{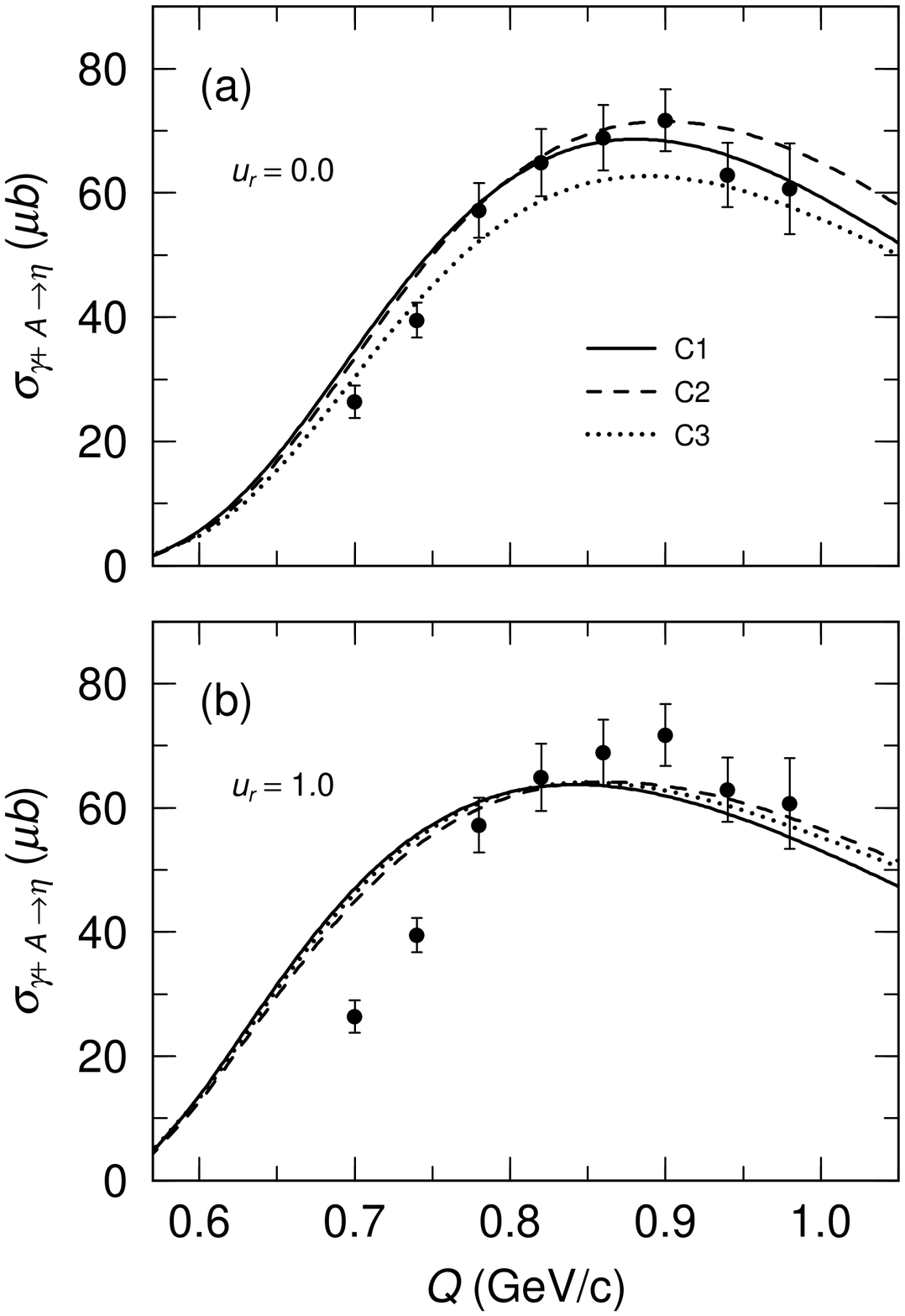}}

\bigskip
\noindent
{{\bf Fig. 5:}
Cross section for eta photo-production with $^{12}$C targets
using  $M^{*}_N / M_N = 0.7$, with (a) $u_r =0$ and (b) $u_r = 1$.
The solid, dashed and dotted curves represent the sets of results 
C1, C2 and C3, whose meanings are explained in the text. 
The filled circles show the experimental data \cite{Yorita}.
}

\end{document}